\documentclass[twocolumn,amsmath,amssymb]{revtex4}

\usepackage{graphicx}

\begin{document}

\title{sscMap: An extensible Java application for connecting small-molecule drugs using gene-expression signatures}
\author{Shu-Dong Zhang}
\affiliation{MRC Toxicology Unit, Hodgkin Building, Lancaster
Road, University of Leicester, Leicester, UK}
\affiliation{Centre for Cancer Research and Cell Biology (CCRCB), Queen's University Belfast,
    Belfast, UK}
\author{Timothy W. Gant}
\affiliation{MRC Toxicology Unit, Hodgkin Building, Lancaster
Road, University of Leicester, Leicester, UK}


\begin{abstract}
Background:
Connectivity mapping is a process to recognize novel pharmacological and toxicological properties in small molecules by comparing their gene expression signatures with others in a database. A simple and robust method for connectivity mapping with increased specificity and sensitivity was recently developed, and its utility demonstrated using experimentally derived gene signatures.

Results:
This paper introduces sscMap (statistically significant connections' map), a Java application designed to undertake connectivity mapping tasks using the recently published method. The software is bundled with a default collection of reference gene-expression profiles based on the publicly available dataset from the Broad Institute Connectivity Map 02, which includes data from over 7000 Affymetrix microarrays, for over 1000 small-molecule compounds, and 6100 treatment instances in 5 human cell lines. In addition, the application allows users to add their custom collections of reference profiles and is applicable to a wide range of other 'omics technologies.

Conclusions:
The utility of sscMap is two fold. First, it serves to make statistically significant connections between a user-supplied gene signature and the 6100 core reference profiles based on the Broad Institute expanded dataset. Second, it allows users to apply the same improved method to custom-built reference profiles which can be added to the database for future referencing. The software can be freely downloaded from http://purl.oclc.org/NET/sscMap.
\end{abstract}


\maketitle

\section{Background}
Interaction of a drug or chemical with a biological system can result in a gene-expression profile or signature characteristic of the event. Lamb et al were the first to propose using these data in
connectivity mapping to make connections between the pharmacological and toxicological properties of small molecules \citep{Lamb-etal-Science-2006}. The three key components in the working of a connectivity map are: 1) a collection of pre-built reference gene-expression profiles that serves
as a core database; 2) a query gene signature, usually prepared by the user, which best characterizes a compound-induced biological state and; 3) a similarity metric to quantify the connection between a gene signature and a reference profile. In a previous publication \citep{Zhang&Gant-BMCBioinformatics2008} we presented a simple and robust method for connecting small-molecule drugs using gene-expression signatures, the utility of which was shown using three experimentally derived gene signatures from independent studies for HDAC inhibitors \citep{Glaser-etal-MolCancerTher-2003}, estrogen \citep{Frasor-etal-CancerRes-2004}, and
immunosuppressive drugs \citep{Horwitz-etal-Circulation-2004}, respectively. Here in this paper we describe sscMap, a Java application that implements the method, and we focus on its utility and extensibility from a user's perspective.

\section{Implementation}
The software was built using Java programming language (Java Platform, Standard Edition 6). JFC/Swing classes were used to provide a Graphical User Interface (GUI) of the program. In designing the software, user extensibility was considered to be an important feature. To this end each individual reference profile is stored as a separate file on the disk. This setting greatly enhances flexibility and extensibility of the software, as it allows users to supplement the default collection of reference profiles, or to build custom collection of reference profiles by following a simple contract specified in the README file. In the execution of the program, a set of reference profiles are first loaded to memory and compared to all the query gene signatures to calculate the related connection scores and p-values, the memory is then released and the program proceeds to load another set of reference profiles from disk. This memory management scheme enables the program to handle an anticipated increasing number of reference profiles at a moderate cost of speed. If all the available reference profiles were residing in memory, the number of reference profiles allowed would soon be limited.

\section{Results}
\subsection{The core database}
The Broad Institute released Build 02 of their Connectivity Map (http://www.broad.mit.edu/cmap/) with an expanded dataset over the 01 version with more compounds utilized. Using these data we constructed 6100 reference gene-expression profiles using the method described in our previous report \citep{Zhang&Gant-BMCBioinformatics2008}. In brief the genes were primarily sorted by the absolute value of log-ratios in descending order, so that the most differentially expressed gene has the highest rank. Thus sscMap comes with a default core database of 6100 reference profiles, each characterizes a treatment instance as described in \citep{Lamb-etal-Science-2006}. This core database covers the treatment instances of over 1000 small-molecule compounds applied to 5 human cell lines. So the primary utility of sscMap is for users who want to compare their gene signatures to the reference profiles based on the Connectivity Map 02 dataset. The benefits of the method implemented in this application include a more principled statistical procedure \citep{Tian-etal-PNAS-2005,Efron-Tibshirani-AnnApplStat-2007, Chen-etal-Bioinformatics-2007}, effective safeguards against false connections, and an increased sensitivity. The sscMap program can be run in two execution modes: as a command line program, or as a GUI (Graphical User Interface) application. In the simple command line mode using the built-in core database users can simply put their gene signature files into the queries folder and run the application. Detailed instructions and guided tours on how to run the program in GUI mode can be found in the accompanying README file.  For more advanced application the database can be added to at the users discretion as described below.

\subsection{Extensibility: Building custom extensions}
Users who want a greater capacity for comparison than the built-in database can build up their own custom reference profiles and apply the same scoring scheme and statistical testing procedures introduced in \citep{Zhang&Gant-BMCBioinformatics2008}. This section briefly describes how users can customize the application.

A plain text file, {\em parameters.ini} for the command line mode, or {\em for-gui/default-parameters.ini} for the GUI mode, sets the key parameters used by the program, e.g., where to find the ref-files (reference gene-expression profiles). In the default settings, we specified  {\em reffiles} as the default folder, where the 6100 reference profiles are stored. It is possible to supplement them by simply putting more similarly built ref-files into the default folder. Users can also create a new directory, for example, {\em custom-reffiles}, and put all the custom ref-files there, and then point the reference profiles folder to that directory, either by editing the {\em parameters.ini} file for the command line, or by browsing to the custom directory in the GUI mode.

As an example, we have included with sscMap a folder {\em custom-example}, which contains all the key components of a customized extension to the application. Following the example provided users should be able to build their own extension. A more detailed description of the general contracts for adding a custom collection of reference profiles to sscMap can be found in the README file accompanying the software.

\subsection{Flexibility: Treatment set definition}
The sscMap software downloads with a default ref-files folder containing 6100 pre-built reference expression profiles. An example ref-file is {\em azathioprine\_0.1mM\_MCF7\_338.ref.tab}, which characterizes the biological state of MCF7 cells treated with 0.1mM azathioprine.
The name of a typical ref-file is divided by the underscore character \_, the default field separator, into 4 fields: drug name, dose, cell type, and instance ID, respectively. The program allows users to specify which field(s) to use for defining a ``treatment set" (A term we use interchangeably with ``reference set", or simply ref-set elsewhere). Our preferred choice for the default ref-files is to use the 3 fields: drug name (field 0), dose (field 1), and cell type (field 2) together to define a treatment set, meaning that only reference profiles with the same drug, same dose, and same cell type should be regarded as forming a set in the set-level analysis. The original Connectivity Map uses only the drug name to define a treatment set, disregarding possible difference in dose and cell type. This tends to average out the distinct characteristics attributable to the cell type or dose difference, making some set-level connections insignificant or their interpretation difficult. We described in the discussion section of \citep{Zhang&Gant-BMCBioinformatics2008} why it was preferable to use 3 fields to define a treatment set. However, the program does not force users to follow this preference. With sscMap, users can choose whatever field(s) they feel appropriate to define a set. One extreme case is to use all the fields of a ref-file name, and consequently each treatment set will have only one treatment instance (such a treatment set is called a singleton set) and this reduces to the instance-level analysis.

In the {\em custom-example} folder, the custom ref-files names are divided by a custom filed separator, -\/-, ie, two hyphen characters, into 4 fields: drug name, dose, tissue type, and time point, as in {\em Drug2-\/-LowDose-\/-Tissue3-\/-Day11.ref.tab}. Treatment sets are defined using two fields, Drug name (field 0) and Tissue type (field 2) in this case. Thus the example here demonstrates the flexibility offered by the application: users have the freedom to choose their own field separator, the number of fields, and which fields to define a treatment set.

\subsection{Example 1: Using the default core collection of ref-files}
As an example of querying the default core database, we used the five gene signatures previously reported in \citep{Zhang&Gant-BMCBioinformatics2008}. Note that this default core database contains 6100 reference gene-expression profiles, which is a much-expanded collection as compared to 453 in \citep{Zhang&Gant-BMCBioinformatics2008}. The connection scores and p-values for the Estrogen gene signature are shown here in Figure \ref{fig-Estrogen-SSC-plot}
in graphical view. The detailed tabulated results can be found as a tab file ({\em Estrogen.sig.sscmap.tab}) in the {\em results} folder within the downloaded software.

\subsection{Example 2: Using a custom collection of ref-files}
In the folder {\em custom-example} we provided a small collection of 18 custom reference profiles, constructed using Affymetrix RAT230\_2 microarray data. We then queried this small database of custom reference profiles using 2 specially prepared gene signatures based on mouse cDNA microarray data. To query the rat reference profiles using mouse gene signatures, we first converted the gene IDs on the mouse array to the Affymetrix Rat230\_2 probeset IDs, using the annotation file provided by Affymetrix. The biological contexts of these reference profiles and gene signatures in this example are not directly relevant, as we are here simply demonstrating the possibility of extending the sscMap software with custom reference profiles. In Table \ref{tab-mouse},
we list all the connections of the 6 reference sets, each containing 3 individual reference profiles, to one of the mouse gene signatures.

\begin{figure}
\caption{A screenshot of the sscMap program displaying the volcano plot for the Estrogen gene signature. The x-axis is for the standardized connection score, while the y-axis is for $-\log _{10} p$. The green horizontal line is for the pre-set threshold p-value. Any data points above that line are considered as statistically significant. In the example shown in this figure, the threshold p-value was set as $1/N=1/3738$.}
\label{fig-Estrogen-SSC-plot}
\end{figure}

\begin{table}
\caption{\label{tab-mouse}   The connections of the rat reference profiles with a mouse gene signature. n, set size; s, set score; p, p value; $\sigma$, the standard deviation of random scores; $z=s/\sigma$.}
\begin{center}
\begin{tabular}{cccccc}
\hline
ref-set&n&s&p&$\sigma$&z\\
\hline
Drug2-\/-Tissue2&3&0.0036&0.9716&0.0908&0.0394\\
Drug1-\/-Tissue1&3&0.1553&0.0468&0.0790&1.9651\\
Drug1-\/-Tissue2&3&0.0167&0.8638&0.0983&0.1694\\
Drug1-\/-Tissue3&3&0.0221&0.7816&0.0799&0.2767\\
Drug2-\/-Tissue1&3&-0.1934&0.0574&0.1020&-1.8953\\
Drug2-\/-Tissue3&3&0.0648&0.3674&0.0709&0.9143\\
\hline
\end{tabular}
\end{center}
\end{table}

\section{Conclusions}
The utility of sscMap is two fold. First, it serves to make statistically significant connections between a user-supplied gene signature and the 6100 core reference profiles based on the Broad Institute expanded dataset. Second, it allows users to apply the scoring scheme and statistical procedures described in \citep{Zhang&Gant-BMCBioinformatics2008} to custom-built reference profiles which can be added to the database for future referencing.

\section{Availability and requirements}

Project name: sscMap

Project home page: http://purl.oclc.org/NET/sscMap

Operating system(s): Platform independent

Programming language: Java

Other requirements: Java Runtime Environment 1.6 or later version is required to run the program.

License: None required for research and academic use.

Any restrictions to use by non-academics: For commercial use, please contact the authors.

\section{Authors' contributions}
SDZ and TWG designed the study. SDZ developed the algorithm,
implemented the method, and analyzed the data. SDZ and TWG wrote
the paper. All authors read and approved the final manuscript.


\section{Acknowledgments}
\noindent We thank the reviewers for their constructive comments and suggestions.  Financial support for this project was provided by the Medical Research Council UK (MRC) and the work carried out with the support of all members of the Systems Toxicology Group of the MRC Toxicology Unit.  SDZ thanks Qing Wen for helpful discussions on a searching algorithm in the implementation of the application.



\end{document}